\documentclass[twocolumn,showpacs,amsmath,amssymb,amsfonts,floatfix,prl,aps]{revtex4}
\usepackage{graphicx}
\usepackage{color}

\begin{document}
\title{Doping induced metal-insulator phase transition in NiO}
\author{Y. Shinohara$^1$}
\author{S. Sharma$^{1,2}$}
\email{sharma@mpi-halle.mpg.de}
\author{J. K. Dewhurst$^1$}
\author{S. Shallcross$^3$}
\author{N. N. Lathiotakis$^{1,4}$}
\author{E. K. U. Gross$^1$}
\affiliation{1 Max-Planck-Institut f\"ur Mikrostrukturphysik, Weinberg 2, 
D-06120 Halle, Germany.}
\affiliation{2. Department of Physics, Indian Institute of Technology, Roorkee, 247667, Uttarkhand, India.}
\affiliation{3. Lehrstuhl f\"ur Theoretische Festkorperphysik, Staudstr. 7-B2 91058 Erlangen, Germany.}
\affiliation{4. Theoretical and Physical Chemistry Institute, National Hellenic Research Foundation,
Vass. Constantinou 48, GR-11635 Athens, Greece}

\date{\today}

\begin{abstract}
The insulator to metal phase transition in NiO is studied within the framework of
reduced density matrix functional theory and density functional theory. We find that the
spectral density obtained using reduced density matrix functional theory
is in good agreement with experiments both undoped as well as doped NiO.
We find that the physical description of the hole-doping induced phase transition \emph{qualitatively
differs} depending on whether NiO is calculated within density functional theory or 
reduced density matrix functional. In the former case the underlying mechanism of the phase transition is 
identified to be a rigid shift of chemical potential, while in the latter case a redistribution 
of the spectral weight drives the transition. These latter results
are found to be in good agreement with both experiments and previous many-body calculations.
\end{abstract}

\maketitle

%%%%%%%%%%%%%%
% Introduction
%%%%%%%%%%%%%%

The anti ferromagnetic (AFM) Mott insulators present one of the outstanding challenges to the first principles
treatment of extended solids. The prototypical example of such a system is NiO, an AFM 
Mott insulator with a measured gap of 4.1eV and a magnetic moment of 1.7$\mu_B$. 
The standard local density approximation (LDA) within density functional theory (DFT) predicts a metallic spectrum for 
NiO, in fundamental disagreement with experimental reality. The inclusion of spin polarization via the local
spin density approximation (LSDA) results in a small Kohn-Sham(KS) gap, and a description of NiO as a Slater insulator.
However, both the gap and the magnetic moment are severely 
underestimated suggesting that the Slater anti ferromagnetic state obtained within the LSDA
does not describe the true nature of NiO. 

In order to over come this deficiency of the Kohn-Sham (KS) spectra obtained using the LSDA, R\"odl \emph{et al.} proposed
the use of two separate fitting parameters: an on-site Coulomb term $U$ and a scissors
shift $\Delta$ by which the conduction bands are rigidly shifted. $\Delta$ is the difference between the experimental gap and the KS gap obtained
using LSDA+$U$ functional\cite{rodl}. With a certain choice of these two parameters the KS spectra of 
of NiO can be made to agree with computationally expensive many-body techniques such as dynamical mean
field theory (DMFT)\cite{dmft1,dmft2,kunes-nmat,kunes-nio,nekrasov}, reduced density matrix functional 
theory (RDMFT)\cite{sharma08,sharma}, and the $GW$ method\cite{rodl}. This scissors corrected LSDA+$U$ method
comes under the heading of the so called correlated band theory method.

What makes NiO even more interesting is its behavior as a function of doping: one finds an
insulator to metal phase transition (IMT) on doping the system with Li, which amounts to hole doping\cite{nio-expd}.
The physics of this phase transition was found to be very rich due to a subtle interplay of charge transfer and Mott 
localization: despite being a text book Mott insulator, NiO also has a strong charge transfer character due to 
the large overlap (in energy) of the Ni-$d$ and O-$p$ states. Any theory attempting to capture this IMT in NiO 
must be capable of treating Mott correlations and charge transfer effects at an equal footing\cite{kunes-nio}, 
presenting a significant theoretical challenge. 

In the present work we study the IMT in NiO using three different approaches: the local spin density approximation (LSDA) 
within DFT, correlated band theory method, LSDA+$U$, and a many-body technique, RDMFT. In doing so we demonstrate that
even though at zero doping all these methods give similar spectra, the physics of hole doping induced phase transition 
is \emph{qualitatively different} for the different theoretical methods: within band and correlated band theory methods the
metalization occurs due to rigid shifting of the chemical potential into the valence band, with the separation between 
the Hubbard bands remaining approximately constant. In total contrast to this, within RDMFT one finds that the phase transition is driven by a transfer in spectral weight from the upper and lower Hubbard bands to a low energy peak, known as the correlated peak. These latter results we find to be in good agreement with previous many-body results obtained using DMFT\cite{kunes-nio}.

Recently, RDMFT has shown potential for correctly treating band as well as Mott insulators\cite{sharma08,nek10,sharma}, or in other words, treating both Mott correlations and charge transfer effects equally well. Within RDMFT, the one-body reduced density matrix (1-RDM) is the basic variable \cite{lodwin,gilbert}

\begin{align}\label{1rdm}
 \gamma({\bf x}, {\bf x'})\equiv N\int d^3x_2\ldots d^3x_N
 \Psi({\bf x},{\bf x}_2 \ldots {\bf x}_N)
 \Psi^*({\bf x}',{\bf x}_2 \ldots {\bf x}_N),
\end{align}
where $\Psi$ denotes the many-body wave function and $N$ is the total number of
electrons. Diagonalization of $\gamma$ produces a set of orthonormal Bloch functions, 
the so called natural orbitals\cite{lodwin}, 
$\varphi_{i{\bf k}}$, and occupation numbers, $n_{i{\bf k}}$. 
Extending RDMFT to the truly non-collinear magnetic case\cite{sharma}, by treating the natural orbitals as two component Pauli-spinors,
leads to the spectral representation

\begin{align}
\gamma({\bf x},{\bf x}')=\sum_{i{\bf k}} 
 n_{i{\bf k}}\varphi_{i{\bf k}}({\bf x})\varphi_{i{\bf k}}^*({\bf x}').
\end{align}
The necessary and sufficient conditions for ensemble $N$-representability
of $\gamma$ were provided, in a classic work, by Coleman\cite{coleman}. These conditions require

\begin{align}
&&0\le n_{i{\bf k}}\le 1 \\ \nonumber
&&\sum_{i{\bf k}}n_{i{\bf k}}=N. 
\end{align}

In terms of $\gamma$, the total ground-state energy \cite{gilbert} of the 
interacting system is (atomic units are used throughout)

\begin{align} \label{etot} \nonumber
E[\gamma]=&-\frac{1}{2} {\rm tr}_{\sigma}\int\lim_{{\bf r}\rightarrow{\bf r}'}
\nabla_{\bf r}^2 \gamma({\bf r},{\bf r}')\,d^3r'
+\int\rho({\bf r}) V_{\rm ext}({\bf r})\,d^3r \\
&+\frac{1}{2}  \int 
\frac{\rho({\bf r})\rho({\bf r}')}
{|{\bf r}-{\bf r}'|}\,d^3r\,d^3r'+E_{\rm xc}[\gamma],
\end{align}
where $\rho({\bf r})={\rm tr}_{\sigma}\gamma({\bf r},{\bf r})$, $V_{\rm ext}$ is a given
external potential, and $E_{\rm xc}$ we call the exchange-correlation (xc)
energy functional. In principle, Gilbert's \cite{gilbert} generalization of the
Hohenberg-Kohn theorem to the 1-RDM guarantees the existence of a functional
$E[\gamma]$ whose minimum, for fixed a $V_{\rm ext}$, yields the exact $\gamma$ 
and the exact ground-state energy. In practice, however, the exchange-correlation energy 
is an unknown functional of $\gamma$ and needs to be approximated. While there 
are several known approximations for the xc energy functional, the most promising for extended systems is the
power functional\cite{sharma08} where the xc energy is given by

\begin{align} 
E_{\rm xc}[\gamma]= -\frac{1}{2}\int \, \int d^3r' d^3r
 \frac{|\gamma^{\alpha}({\bf r},{\bf r}')|^2}{|{\bf r}-{\bf r}'|}
\end{align}
with $\alpha$ indicating the power in the operator sense. In view of the universality of the functional $E_{\rm xc}[\gamma]$,
the value of $\alpha$ should, in principle, be system-independent. A few "optimum values" of $\alpha$ have been suggested in the literature\cite{power_finite,sharma08,esa11}; in the present work $\alpha$ is fixed to 0.56. 

In order to the study the doping dependent metalization of NiO one crucially needs spectral information. To obtain this
information from RDMFT, which by its very nature is a ground-state theory, is a difficult task. 
In this work we extract spectral density from RDMFT using the method recently proposed in Ref. [\onlinecite{sharma}].
Within this method the diagonal of the spectral density, i.e. the density of states, is determined using the following relation:

\begin{eqnarray}\label{dos}
{\rm DOS}= 2\pi 
\sum_{\lambda}\left[ n_{i\bf k} \delta(\omega-\epsilon^-_{i\bf k})
+ (1-n_{i\bf k}) \delta(\omega+\epsilon^+_{i\bf k}) \right]
\end{eqnarray}
where

\begin{align}\label{dedn}
 \epsilon^{\pm}_{i\bf k}= 
 \left.\frac{\partial E[\{\phi\},\{n\}]}
 {\partial n_{i\bf k}}\right|_{n_{i{\bf k}=1/2}} - \mu.
\end{align}
where, $\mu$ is the chemical potential. This method is known to produce accurate spectra for finite systems\cite{nek12} as well as solids\cite{sharma}. 
Following the above procedure the spectral density for doped and undoped NiO is calculated using the full-potential 
linearized augmented plane wave \cite{singh} code Elk\cite{elk}.

%%%%%%%%%%%%%%
% NiO doped
%%%%%%%%%%%%%%
\begin{figure}[ht]
\vspace{0.5cm}
\centerline{\includegraphics[width=\columnwidth,angle=-0]{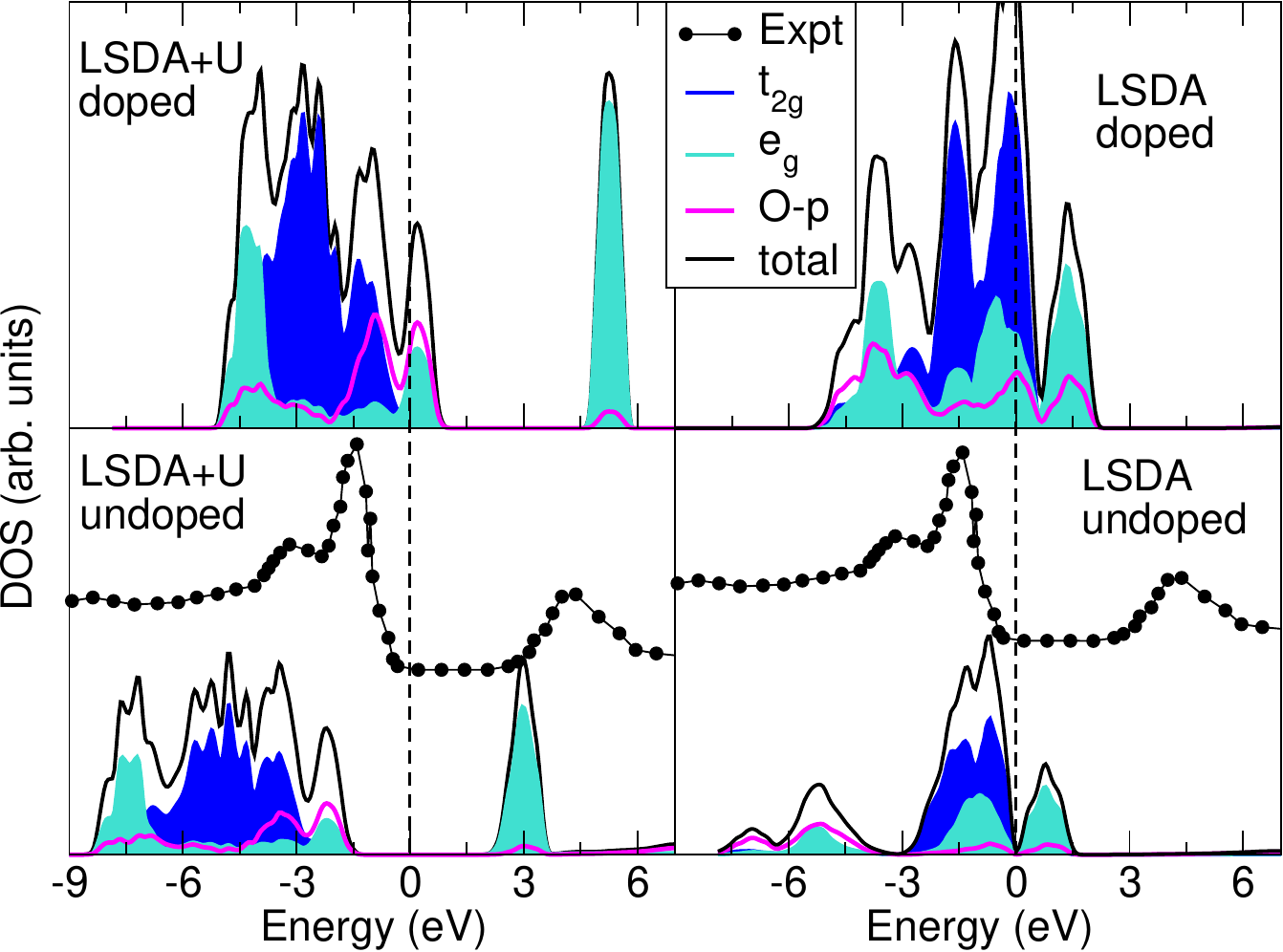}}
\caption{(Color online) Projected and total density of states for NiO as a function of energy (in eV). Results are shown for undoped (lower panels) and hole doped (upper panels) NiO. The results are obtained using two different functionals within DFT; the LSDA (right panels) and the LSDA+$U$ method (left panels).
Experimental data from Ref \onlinecite{nio-expud} is also presented for comparison.}
\vspace{0.7cm}
\label{nio-lda}
\end{figure}

We first examine the behaviour of undoped NiO within band and correlated band theory by using LSDA and LSDA+$U$ methods; 
results for total DOS and the site and symmetry projected spectral density are presented in Fig.~(\ref{nio-lda}). 
One can observe that while the LSDA band gap is grossly underestimated as compared to experiment, a 
\emph{rigid shift} of the valence band to lower energies would result in a spectrum with an 
overall shape in good agreement with experiment. Indeed, the correct ordering of the $t_{2g}$ 
and $e_g$ states is obtained within LSDA, i.e., the band gap, even though underestimated, separates a $t_{2g}$ valence and a $e_g$ conduction band, with a substantial overlap of $e_g$ and $t_{2g}$ states in the valence band.

On applying an on-site Coulomb repulsion the gap opens, but at the cost of a deterioration in the agreement of the overall spectral weight with experiment. In particular, the band ordering is now found to be incorrect at precisely the
value of $U$ that leads to the correct magnitude of the gap, which occurs between purely $e_g$ states. 
This striking defect of the LSDA+$U$ treatment of NiO was also noticed in Ref. \onlinecite{rodl}, and the 
use of a smaller $U$ with an additional external parameter, the so called scissors correction $\Delta$, 
was suggested as a remedy. However, while the spectrum of the equilibrium ground state is improved by this procedure, albeit with the use of an additional fitting parameter, such an approach cannot be used to study the insulator to metal transition in NiO: the very meaning of the parameter $\Delta$ is lost once the material enters the metallic phase.
Upon doping the KS spectrum shows a rather simple behaviour for both LSDA and LSDA+$U$ methods; the chemical potential rigidly moves into the valence band leading in consequence to metalization. In the case of  LSDA+$U$ method this implies same value of $U$ across the phase transition. 
%We may estimate the distribution of the added hole by integrating the site and symmetry projected DOS up to the chemical potential. The values of the electronic charge in various symmetry states are presented in Table I. It is
%clear that in total contrast to the experimental situation \cite{nio-expd}, within the LSDA and LSDA+$U$ methods the added holes are equally distributed between the Ni $d$ and O $p$-states.

%\begin{table}[t]
%\setlength{\tabcolsep}{0.1cm}
%\begin{tabular}{ccc}
%            & Undoped     &  Doped  \\ \hline
%            & LSDA        & \\  \hline
%Ni-$t_{2g}$ & 5.42        &  4.81 \\
%Ni-$e_g$    & 2.43        &  2.43 \\
%O-$p$       & 1.73        &  1.51  \\ \hline \hline
%            & LSDA+$U$    & \\  \hline
%Ni-$t_{2g}$ & 5.52        & 5.52 \\
%Ni-$e_g$    & 2.19        & 1.80 \\
%O-$p$       & 1.78        & 1.51 \\ \hline \hline
%            & RDMFT       & \\  \hline
%Ni-$t_{2g}$ & 5.49        & 5.49 \\
%Ni-$e_g$    & 2.27        & 2.27 \\
%O-$p$       & 1.88        & 1.08 \\\hline \hline
%\end{tabular}
%\caption{Number of electrons in various site and symmetry projected states. Results obtained for undoped and
%hole doped (with 1.2 holes per formula unit) NiO obtained using the LSDA, the LSDA+$U$ method, and RDMFT.}
%\end{table}

\begin{figure}[ht]
\vspace{0.5cm}
\centerline{\includegraphics[width=\columnwidth,angle=-0]{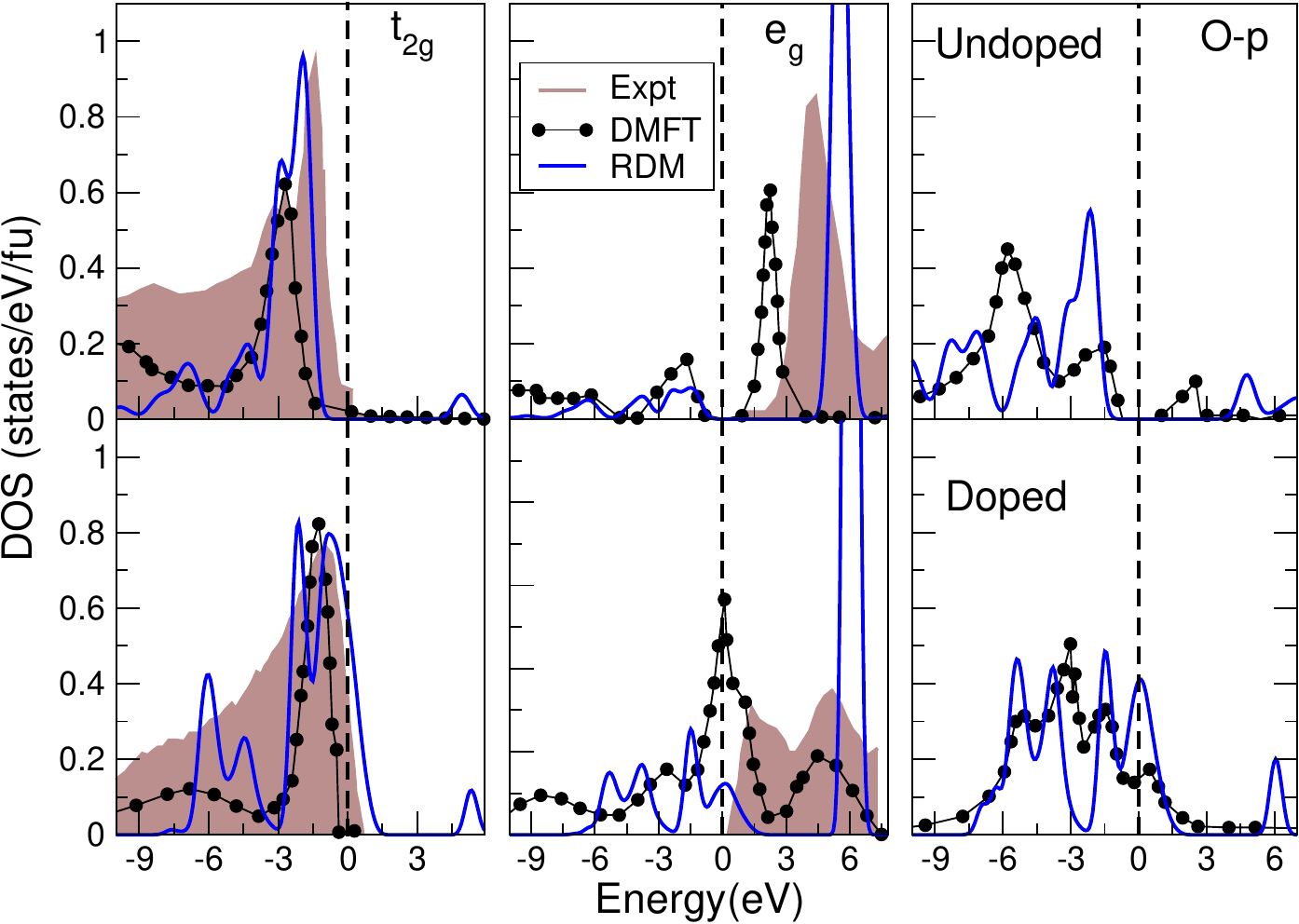}}
\caption{(Color online) Projected density of states (in states/eV/formula-unit) for NiO, obtained using RDMFT, as a function of energy (in eV).
Results are presented for undoped (upper panel), hole doped with 1.4 holes per formula unit (lower panel). 
Experimental results from Ref. \onlinecite{nio-expud} and \onlinecite{nio-expd} and DMFT results from Ref. \onlinecite{nekrasov,kunes-nio} 
are presented for comparison.}
\vspace{0.7cm}
\label{nio-dop}
\end{figure}

% RDMFT
We now consider the situation where NiO is treated within RDMFT, see Fig.~(\ref{nio-dop}). In good agreement with previous 
many-body studies\cite{ren,fuji}, we find that the conduction band is almost entirely  $e_g$ in character. A substantial 
overlap of Ni-$d$ and O-$p$ states may be seen in the valence spectrum, highlighting the presence of 
charge transfer effects in NiO.

For the undoped case, the spectral density is in good agreement with experiments\cite{nio-expud}. The RDMFT results also 
agree very well with previous DMFT calculations\cite{nekrasov,kunes-nio,ren,fuji} with one important difference between 
the two; the experimental data shows a shoulder at 
-3eV, which is well captured by $t_{2g}$ like states within RDMFT but is missing in DMFT results. 
The value of the band-gap obtained using RDMFT (4.88eV) is larger than experiments (4.1eV), and the magnetic moment 
(1.52$\mu_B$) smaller than the experimental value of 1.7$\mu_B$.
There are two reasons for the smaller value of the magnetic moment within RDMFT as compared to experiment. 
Firstly, the calculations are performed with the FP-LAPW method 
in which space is divided into spheres around the atoms, the so called muffin-tins, and an interstitial region. 
In the case of fully non-collinear magnetic calculations the magnetic moment per site is calculated by integrating the 
magnetization vector field inside the muffin-tin. This implies the loss of a
small part of the moment to the interstitial region. Secondly, the power functional induces a slight non-collinearity in 
the magnetization leading to yet more loss in the integrated $z$-projected moment.

Turning to the hole doping of NiO, we find that the effect on the spectral density is strikingly different
within RDMFT, as compared to both the LSDA and LSDA+$U$ methods. Hole doping is found to lead to a redistribution
of spectral weight of the Ni $e_g$ states from the upper and the lower Hubbard bands towards the chemical potential, 
which in turn leads to an IMT. The Ni $t_{2g}$ like states remain almost the same as in undoped case. 
If one uses the correlated band theory definition of $U$ as being roughly equal to the distance between the 
upper Hubbard band and the correlated peak, then it is evident that the value of $U$ changes as a function of doping. 
%An additional striking difference between DFT and RDMFT results is that the added hole is mostly localized in the 
%O $p$-states, in marked contrast to the equal distribution between Ni $d$- and O $p$-states found within LSDA and LSDA+$U$ method. 
These results are in good agreement with experiments as well as previous DMFT data and show marked difference from LSDA and LSDA+$U$
data in that the doping induced DOS in not just a rigid shift of the Fermi energy into the valence band. 

Turning to the quantitative description of the phase transition afforded by RDMFT, we find that, like DMFT, the substantial metalization 
occurs at a much higher value of hole doping (1.2 holes per formula-unit for RDMFT)
than observed in experiments (0.5 holes). There are two principle reasons for this. Firstly, the undoped gap for NiO 
(4.88eV) is larger than the experimental value, and hence additional hole doping with be required to drive the material
to the metallic state. Secondly, we do not study the effect 
of an actual impurity added to the system but rather the hole doping is simulated by the removal of electronic charge from the unit cell while adding a constant compensating background to ensure charge neutrality. This method is commonly known as the virtual crystal approximation.

To summarize we have presented the physics behind the doping induced insulator to metal phase transition in NiO and have
found that the physics of phase transition is brought out in strikingly different ways by different theoretical methods: within DFT based studies metalization occurs due to a rigid shifting of the chemical potential into the valence band, with the separation between the Hubbard bands remaining approximately constant. In contrast, within RDMFT the phase transition 
is driven by a transfer in spectral weight from the upper and lower Hubbard bands to what is known as the correlated 
peak. We thus find that RMDFT treatment of NiO is much close to many-body theories such as the DMFT, than to the LSDA or 
correlated band approaches

%\bibliography{nio}

\begin{thebibliography}{21}
\expandafter\ifx\csname natexlab\endcsname\relax\def\natexlab#1{#1}\fi
\expandafter\ifx\csname bibnamefont\endcsname\relax
  \def\bibnamefont#1{#1}\fi
\expandafter\ifx\csname bibfnamefont\endcsname\relax
  \def\bibfnamefont#1{#1}\fi
\expandafter\ifx\csname citenamefont\endcsname\relax
  \def\citenamefont#1{#1}\fi
\expandafter\ifx\csname url\endcsname\relax
  \def\url#1{\texttt{#1}}\fi
\expandafter\ifx\csname urlprefix\endcsname\relax\def\urlprefix{URL }\fi
\providecommand{\bibinfo}[2]{#2}
\providecommand{\eprint}[2][]{\url{#2}}

\bibitem[{\citenamefont{R\"odl et~al.}(2009)\citenamefont{R\"odl, Fuchs,
  Furthm\"uller, and Bechstedt}}]{rodl}
\bibinfo{author}{\bibfnamefont{C.}~\bibnamefont{R\"odl}},
  \bibinfo{author}{\bibfnamefont{F.}~\bibnamefont{Fuchs}},
  \bibinfo{author}{\bibfnamefont{J.}~\bibnamefont{Furthm\"uller}},
  \bibnamefont{and}
  \bibinfo{author}{\bibfnamefont{F.}~\bibnamefont{Bechstedt}},
  \bibinfo{journal}{Phys. Rev. B} \textbf{\bibinfo{volume}{79}},
  \bibinfo{pages}{235114} (\bibinfo{year}{2009}).

\bibitem[{\citenamefont{Metzner and Vollhardt}(1989)}]{dmft1}
\bibinfo{author}{\bibfnamefont{W.}~\bibnamefont{Metzner}} \bibnamefont{and}
  \bibinfo{author}{\bibfnamefont{D.}~\bibnamefont{Vollhardt}},
  \bibinfo{journal}{Phys. Rev. Lett.} \textbf{\bibinfo{volume}{62}},
  \bibinfo{pages}{324} (\bibinfo{year}{1989}).

\bibitem[{\citenamefont{Georges et~al.}(1996)\citenamefont{Georges, Kotliar,
  Krauth, and Rozenberg}}]{dmft2}
\bibinfo{author}{\bibfnamefont{A.}~\bibnamefont{Georges}},
  \bibinfo{author}{\bibfnamefont{G.}~\bibnamefont{Kotliar}},
  \bibinfo{author}{\bibfnamefont{W.}~\bibnamefont{Krauth}}, \bibnamefont{and}
  \bibinfo{author}{\bibfnamefont{M.~J.} \bibnamefont{Rozenberg}},
  \bibinfo{journal}{Phys. Rev. B} \textbf{\bibinfo{volume}{68}},
  \bibinfo{pages}{13} (\bibinfo{year}{1996}).

\bibitem[{\citenamefont{Kunes et~al.}(2008)\citenamefont{Kunes, Lukoyanov,
  Anisimov, Scalettar, and Pickett}}]{kunes-nmat}
\bibinfo{author}{\bibfnamefont{J.}~\bibnamefont{Kunes}},
  \bibinfo{author}{\bibfnamefont{A.~V.} \bibnamefont{Lukoyanov}},
  \bibinfo{author}{\bibfnamefont{V.~I.} \bibnamefont{Anisimov}},
  \bibinfo{author}{\bibfnamefont{R.~T.} \bibnamefont{Scalettar}},
  \bibnamefont{and} \bibinfo{author}{\bibfnamefont{W.~E.}
  \bibnamefont{Pickett}}, \bibinfo{journal}{Nat. Mat.}
  \textbf{\bibinfo{volume}{7}}, \bibinfo{pages}{198} (\bibinfo{year}{2008}).

\bibitem[{\citenamefont{Kunes et~al.}(2007)\citenamefont{Kunes, anisimov,
  Lukoyanov, and Vollhardt}}]{kunes-nio}
\bibinfo{author}{\bibfnamefont{J.}~\bibnamefont{Kunes}},
  \bibinfo{author}{\bibfnamefont{V.~I.} \bibnamefont{anisimov}},
  \bibinfo{author}{\bibfnamefont{A.~V.} \bibnamefont{Lukoyanov}},
  \bibnamefont{and}
  \bibinfo{author}{\bibfnamefont{D.}~\bibnamefont{Vollhardt}},
  \bibinfo{journal}{Phys. Rev. B} \textbf{\bibinfo{volume}{75}},
  \bibinfo{pages}{165115} (\bibinfo{year}{2007}).

\bibitem[{\citenamefont{Nekrasov et~al.}(2013)\citenamefont{Nekrasov, Pavlov,
  and Sadovskii}}]{nekrasov}
\bibinfo{author}{\bibfnamefont{I.~A.} \bibnamefont{Nekrasov}},
  \bibinfo{author}{\bibfnamefont{N.~S.} \bibnamefont{Pavlov}},
  \bibnamefont{and} \bibinfo{author}{\bibfnamefont{M.~V.}
  \bibnamefont{Sadovskii}}, \bibinfo{journal}{J. Expt and Theor. Phys.}
  \textbf{\bibinfo{volume}{116}}, \bibinfo{pages}{620} (\bibinfo{year}{2013}).

\bibitem[{\citenamefont{Sharma et~al.}(2008)\citenamefont{Sharma, Dewhurst,
  Lathiotakis, and Gross}}]{sharma08}
\bibinfo{author}{\bibfnamefont{S.}~\bibnamefont{Sharma}},
  \bibinfo{author}{\bibfnamefont{J.~K.} \bibnamefont{Dewhurst}},
  \bibinfo{author}{\bibfnamefont{N.~N.} \bibnamefont{Lathiotakis}},
  \bibnamefont{and} \bibinfo{author}{\bibfnamefont{E.~K.~U.}
  \bibnamefont{Gross}}, \bibinfo{journal}{Phys. Rev. B}
  \textbf{\bibinfo{volume}{78}}, \bibinfo{pages}{201103}
  (\bibinfo{year}{2008}).

\bibitem[{\citenamefont{Sharma et~al.}(2009)\citenamefont{Sharma, Shallcross,
  Dewhurst, and Gross}}]{sharma}
\bibinfo{author}{\bibfnamefont{S.}~\bibnamefont{Sharma}},
  \bibinfo{author}{\bibfnamefont{S.}~\bibnamefont{Shallcross}},
  \bibinfo{author}{\bibfnamefont{J.~K.} \bibnamefont{Dewhurst}},
  \bibnamefont{and} \bibinfo{author}{\bibfnamefont{E.~K.~U.}
  \bibnamefont{Gross}}, \bibinfo{journal}{http://www.arxiv.org, cond-mat:
  arXiv:0912.1118}  (\bibinfo{year}{2009}).

\bibitem[{\citenamefont{Elp et~al.}(1992)\citenamefont{Elp, Eskes, Kuiper, and
  Sawatzky}}]{nio-expd}
\bibinfo{author}{\bibfnamefont{J.~V.} \bibnamefont{Elp}},
  \bibinfo{author}{\bibfnamefont{H.}~\bibnamefont{Eskes}},
  \bibinfo{author}{\bibfnamefont{P.}~\bibnamefont{Kuiper}}, \bibnamefont{and}
  \bibinfo{author}{\bibfnamefont{G.~A.} \bibnamefont{Sawatzky}},
  \bibinfo{journal}{Phys. Rev. B} \textbf{\bibinfo{volume}{45}},
  \bibinfo{pages}{1612} (\bibinfo{year}{1992}).

\bibitem[{\citenamefont{Lathiotakis et~al.}(2010)\citenamefont{Lathiotakis,
  Sharma, Helbig, Dewhurst, Marques, Eich, Baldsiefen, Zacarias, and
  Gross}}]{nek10}
\bibinfo{author}{\bibfnamefont{N.~N.} \bibnamefont{Lathiotakis}},
  \bibinfo{author}{\bibfnamefont{S.}~\bibnamefont{Sharma}},
  \bibinfo{author}{\bibfnamefont{N.}~\bibnamefont{Helbig}},
  \bibinfo{author}{\bibfnamefont{J.~K.} \bibnamefont{Dewhurst}},
  \bibinfo{author}{\bibfnamefont{M.~A.~L.} \bibnamefont{Marques}},
  \bibinfo{author}{\bibfnamefont{F.}~\bibnamefont{Eich}},
  \bibinfo{author}{\bibfnamefont{T.}~\bibnamefont{Baldsiefen}},
  \bibinfo{author}{\bibfnamefont{A.}~\bibnamefont{Zacarias}}, \bibnamefont{and}
  \bibinfo{author}{\bibfnamefont{E.~K.~U.} \bibnamefont{Gross}},
  \bibinfo{journal}{Zeitschrift f\"ur Physikalische Chemie}
  \textbf{\bibinfo{volume}{224}}, \bibinfo{pages}{467} (\bibinfo{year}{2010}).

\bibitem[{\citenamefont{L\"odwin}(1955)}]{lodwin}
\bibinfo{author}{\bibfnamefont{P.~O.} \bibnamefont{L\"odwin}},
  \bibinfo{journal}{Phys. Rev. B} \textbf{\bibinfo{volume}{97}},
  \bibinfo{pages}{1974} (\bibinfo{year}{1955}).

\bibitem[{\citenamefont{Gilbert}(1975)}]{gilbert}
\bibinfo{author}{\bibfnamefont{T.~L.} \bibnamefont{Gilbert}},
  \bibinfo{journal}{Phys. Rev. B} \textbf{\bibinfo{volume}{12}},
  \bibinfo{pages}{2111} (\bibinfo{year}{1975}).

\bibitem[{\citenamefont{Coleman}(1963)}]{coleman}
\bibinfo{author}{\bibfnamefont{A.}~\bibnamefont{Coleman}},
  \bibinfo{journal}{Rev. Mod. Phys.} \textbf{\bibinfo{volume}{35}},
  \bibinfo{pages}{668} (\bibinfo{year}{1963}).

\bibitem[{\citenamefont{Lathiotakis et~al.}(2009)\citenamefont{Lathiotakis,
  Sharma, Dewhurst, Eich, Marques, and Gross}}]{power_finite}
\bibinfo{author}{\bibfnamefont{N.}~\bibnamefont{Lathiotakis}},
  \bibinfo{author}{\bibfnamefont{S.}~\bibnamefont{Sharma}},
  \bibinfo{author}{\bibfnamefont{J.}~\bibnamefont{Dewhurst}},
  \bibinfo{author}{\bibfnamefont{F.}~\bibnamefont{Eich}},
  \bibinfo{author}{\bibfnamefont{M.}~\bibnamefont{Marques}}, \bibnamefont{and}
  \bibinfo{author}{\bibfnamefont{E.}~\bibnamefont{Gross}},
  \bibinfo{journal}{Phys. Rev. A} \textbf{\bibinfo{volume}{79}},
  \bibinfo{pages}{040501} (\bibinfo{year}{2009}).

\bibitem[{\citenamefont{Putaja and Rasanen}(2011)}]{esa11}
\bibinfo{author}{\bibfnamefont{A.}~\bibnamefont{Putaja}} \bibnamefont{and}
  \bibinfo{author}{\bibfnamefont{E.}~\bibnamefont{Rasanen}},
  \bibinfo{journal}{Phys. Rev. B} \textbf{\bibinfo{volume}{84}},
  \bibinfo{pages}{035104} (\bibinfo{year}{2011}).

\bibitem[{\citenamefont{Zarkadoula et~al.}(2012)\citenamefont{Zarkadoula,
  Sharma, Dewhurst, Gross, and Lathiotakis}}]{nek12}
\bibinfo{author}{\bibfnamefont{E.~N.} \bibnamefont{Zarkadoula}},
  \bibinfo{author}{\bibfnamefont{S.}~\bibnamefont{Sharma}},
  \bibinfo{author}{\bibfnamefont{J.~K.} \bibnamefont{Dewhurst}},
  \bibinfo{author}{\bibfnamefont{E.~K.~U.} \bibnamefont{Gross}},
  \bibnamefont{and} \bibinfo{author}{\bibfnamefont{N.~N.}
  \bibnamefont{Lathiotakis}}, \bibinfo{journal}{Phys. Rev. A}
  \textbf{\bibinfo{volume}{85}}, \bibinfo{pages}{032504}
  (\bibinfo{year}{2012}).

\bibitem[{\citenamefont{Singh}(1994)}]{singh}
\bibinfo{author}{\bibfnamefont{D.~J.} \bibnamefont{Singh}},
  \emph{\bibinfo{title}{\rm {Planewaves Pseudopotentials and the LAPW Method},
  {Kluwer Academic Publishers, Boston}}} (\bibinfo{year}{1994}).

\bibitem[{elk(2004)}]{elk}
 (\bibinfo{year}{2004}), \urlprefix\url{http://elk.sourceforge.net}.

\bibitem[{\citenamefont{Sawatzky and j.~W.~Allen}(1984)}]{nio-expud}
\bibinfo{author}{\bibfnamefont{G.~A.} \bibnamefont{Sawatzky}} \bibnamefont{and}
  \bibinfo{author}{\bibnamefont{j.~W.~Allen}}, \bibinfo{journal}{Phys. Rev.
  Lett.} \textbf{\bibinfo{volume}{53}}, \bibinfo{pages}{2339}
  (\bibinfo{year}{1984}).

\bibitem[{\citenamefont{Ren et~al.}(2006)\citenamefont{Ren, Leonov, Keller,
  Kollar, Nekrasov, and Vollhardt}}]{ren}
\bibinfo{author}{\bibfnamefont{X.}~\bibnamefont{Ren}},
  \bibinfo{author}{\bibfnamefont{I.}~\bibnamefont{Leonov}},
  \bibinfo{author}{\bibfnamefont{G.}~\bibnamefont{Keller}},
  \bibinfo{author}{\bibfnamefont{M.}~\bibnamefont{Kollar}},
  \bibinfo{author}{\bibfnamefont{I.}~\bibnamefont{Nekrasov}}, \bibnamefont{and}
  \bibinfo{author}{\bibfnamefont{D.}~\bibnamefont{Vollhardt}},
  \bibinfo{journal}{Phys. Rev. B} \textbf{\bibinfo{volume}{74}},
  \bibinfo{pages}{195114} (\bibinfo{year}{2006}).

\bibitem[{\citenamefont{Fujimori et~al.}(1984)\citenamefont{Fujimori, Minami,
  and Sugano}}]{fuji}
\bibinfo{author}{\bibfnamefont{A.}~\bibnamefont{Fujimori}},
  \bibinfo{author}{\bibfnamefont{F.}~\bibnamefont{Minami}}, \bibnamefont{and}
  \bibinfo{author}{\bibfnamefont{S.}~\bibnamefont{Sugano}},
  \bibinfo{journal}{Phys. Rev. B} \textbf{\bibinfo{volume}{29}},
  \bibinfo{pages}{5225} (\bibinfo{year}{1984}).

\end{thebibliography}
\end{document}